\documentclass[aps,prb,twocolumn,groupedaddress,floatfix]{revtex4}

\usepackage{graphicx}

\begin{document}

% Use the \preprint command to place your local institutional report
% number in the upper righthand corner of the title page in preprint mode.
% Multiple \preprint commands are allowed.
% Use the 'preprintnumbers' class option to override journal defaults
% to display numbers if necessary
%\preprint{}

%Title of paper
\title{Gate control of a quantum dot single-electron spin in realistic
confining potentials: anisotropy effects}

% repeat the \author .. \affiliation  etc. as needed
% \email, \thanks, \homepage, \altaffiliation all apply to the current
% author. Explanatory text should go in the []'s, actual e-mail
% address or url should go in the {}'s for \email and \homepage.
% Please use the appropriate macro foreach each type of information

% \affiliation command applies to all authors since the last
% \affiliation command. The \affiliation command should follow the
% other information
% \affiliation can be followed by \email, \homepage, \thanks as well.

\author{Sanjay Prabhakar and James E. Raynolds}
%\email[]{Your e-mail address}
%\homepage[]{Your web page}
%\thanks{}
%\altaffiliation{}
\affiliation{College of Nanoscale Science and Engineering, University
at Albany, State University of New York, Albany, NY}

%Collaboration name if desired (requires use of superscriptaddress
%option in \documentclass). \noaffiliation is required (may also be
%used with the \author command).
%\collaboration can be followed by \email, \homepage, \thanks as well.
%\collaboration{}
%\noaffiliation

\date{\today}

\begin{abstract}
Among recent proposals for next-generation, non-charge-based logic 
is the notion that a single electron can be trapped and its spin can 
be manipulated through the application of gate potentials.
In this paper, we present numerical simulations of such spins in single 
electron devices for realistic (asymmetric) confining potentials in 
two-dimensional electrostatically confined quantum dots.  Using analytical 
and numerical techniques we show that breaking the in-plane rotational 
symmetry of the confining potential leads to a significant effect on the 
tunability of the g-factor with applied gate potentials.  In particular,
anisotropy extends the range of tunability to larger quantum dots.
\end{abstract}

% insert suggested PACS numbers in braces on next line
\pacs{}
% insert suggested keywords - APS authors don't need to do this
%\keywords{}

%\maketitle must follow title, authors, abstract, \pacs, and \keywords
\maketitle

% body of paper here - Use proper section commands
% References should be done using the \cite, \ref, and \label commands
% Put \label in argument of \section for cross-referencing
%\section{\label{}}

\section{Introduction}

The notion of using single-electron spins for quantum computing and 
next-generation logic is an attractive idea that has received considerable
attention in recent 
years.~\cite{loss97,engel04,coish07,awschalom02,hanson07,zutic04,bandyopadhyay00,ranjan08,calarco03,hanson08,biolatti02,young02,thorwart01}

In order to integrate new concepts with existing semiconductor 
technology, a number of researchers have recently explored the 
possibility of using {\em electric fields} generated by imposed gate 
potentials to manipulate single-electron spins in quantum dot 
devices.~\cite{dassarma03,rashba03,rashba03a,laird07,tang05,pingenot08,jiang01,shim08,nowack07,efros01,tarucha08,amasha08,tang06,ortner05,bester05,hoegele05,levitov03,
governale03,nakaoka07,kato08,doty06}

The goal of the present work is to utilize state-of-the-art numerical
techniques to explore the fundamental physics of single-electron spin
devices and to provide realistic information for the practical design of 
such systems. We utilize a finite-element based numerical technique
to study electrostatically defined quantum dots that is similar
to other recently published work.~\cite{bednarek08} 

A key result of the present work is the discovery that spatial symmetry
breaking~\cite{kwasniowski08,konemann05,destefani05} resulting from the 
anisotropy of realistic confining potentials results in a significant 
enhancement of the electric-field tunability of the electron $g$-factor 
over that found for symmetric potentials.

\section{Computational method}

We utilize a multi-scale, multi-physics simulation strategy based on the 
finite element method~\cite{comsol} to provide a realistic description of the 
physics of single-spin devices in three-dimensional geometries.  The ideal is
to solve self-consistently the Maxwell equations of electrostatics 
with the Schroedinger equation in three dimensional geometry.
Unfortunately such a solution is not feasible given currently available 
techniques due to the disparity of length scales in the problem.  We thus 
seek an approximate solution that is built up in stages.

\begin{figure}
\includegraphics[height=6cm]{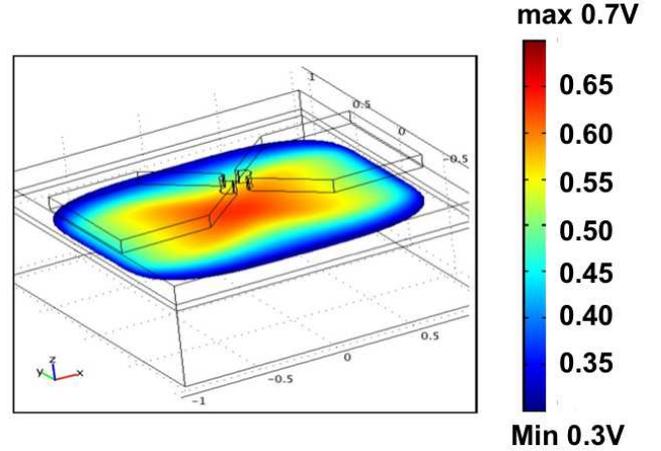}
\caption{\label{figa}(color online) Electrostatic potential for a prototype
single-electron device plotted in the 2DEG layer.  This figure illustrates
a single-spin device consisting of two triangular gates above a 2DEG.
The gates were held at $1V$ and the 2DEG was held at $0V$.  For simplicity
of the electrostatic calculation, the 2DEG was treated as a classical
perfect conductor.  The dimensions of the device in the $x$ and $y$ directions
are $2.8\mu m$ and $1.8\mu m$ respectively and the thickness is $1\mu m$. 
}
\end{figure}

\begin{figure}
\includegraphics[height=6cm]{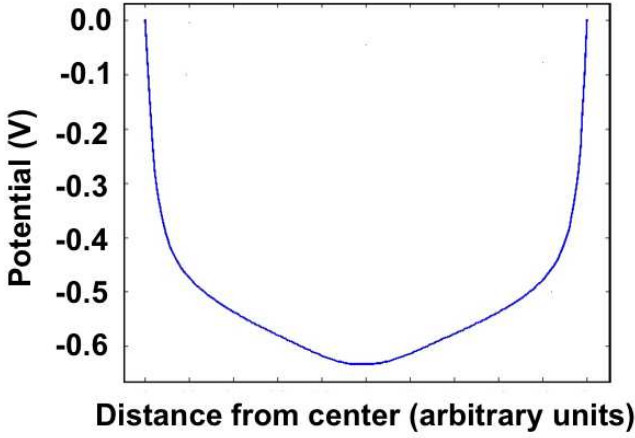}
\caption{\label{figb}Electrostatic confining potential in the 2DEG along 
the symmetry axis of a prototype single-electron device.  This figure was
made by plotting the potential of Fig.~\ref{figa} along a line in the 
2DEG through the symmetry axis of the device (the x-axis of 
Fig.~\ref{figa}, i.e. a line running from one gate to the other) intersecting 
with the central region.}
\end{figure}

\begin{figure}
\includegraphics[height=6cm]{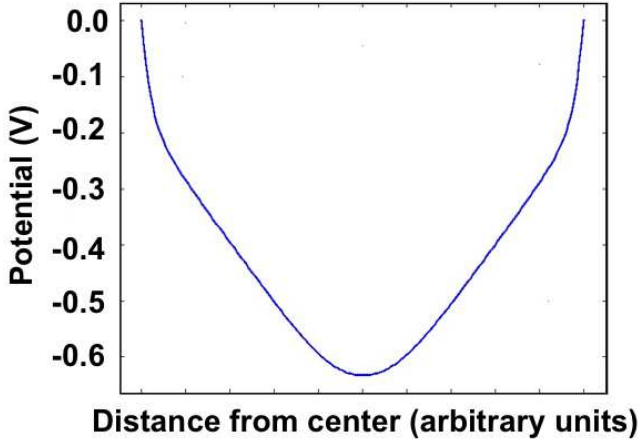}
\caption{\label{figc}Electrostatic confining potential in the 2DEG normal 
to the symmetry axis of a prototype single-electron device.  This figure was
made by plotting the potential of Fig.~\ref{figa} along a line in the 
2DEG normal to the symmetry axis of the device (the y-axis of 
Fig.~\ref{figa}) and intersecting the central region.}
\end{figure}

In the first step of our approach, we construct a three-dimensional 
model of the device and calculate the gate-induced electrostatic potentials 
that cause the formation of a quantum dot in the two-dimensional electron
gas (2DEG) at a AlGaAs/GaAs heterojunction as illustrated in Figs.~\ref{figa},
~\ref{figb}, and~\ref{figc}.  This geometry corresponds to prototype devices
that are under consideration by experimentalists at the University at Albany,
State University of New York.  

In order to obtain the electrostatic solution
for the confining potential, we approximate the 2DEG as a classical perfect
conductor and give it a finite width.  The width ($\approx 0.05\mu m$) is 
unrealistically large from a quantum perspective but is assumed to give a 
reasonable description of the spatial variation of the potential in the 
layer of the 2DEG.  In a subsequent step we treat the 2DEG from a 
realistic quantum mechanical perspective.~\cite{sttern84}

Figs.~\ref{figb} and~\ref{figc} are
one-dimensional plots obtained from Fig.~\ref{figa} by plotting the potential
along a line in the 2DEG along high symmetry directions.  These one-dimensional
potentials are then fit to polynomial forms $P_x(x)$ and $P_y(y)$.  These
are then used as a potential of the form
\begin{equation}
V_{real}(x,y) \approx P_x(x) + P_y(y)
\label{separable}
\end{equation}
to approximate the confining potential of the electron in the Schroedinger
equation.  Before considering electron motion in the above potential,
$V_{real}$, we consider the simpler quadratic potential
\begin{equation}
V_{quad} \equiv {\frac {1}{2}} m \omega_o^2 (\alpha x^2 + \beta y^2)
\label{quadpot}
\end{equation}
that allows for systematic studies.  For convenience we have written the 
strength of the potential in harmonic oscillator form by defining the  
pre-factor ${\frac {1}{2}} m \omega_o^2$.

\begin{figure}
\includegraphics[height=7cm]{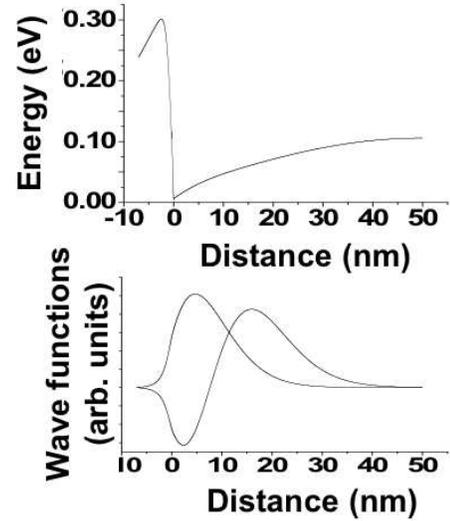}
\caption{\label{figd}Heterojunction self-consistent potential and lowest two 
wave functions of the 2DEG. These results demonstrate consistency of our
results with other published work.~\cite{sttern84}}
\end{figure}

In the second step we calculate the wave functions and self-consistent 
potential at the heterojunction between AlGaAs and GaAs that describes the
formation of the two dimensional electron gas as illustrated in Fig.~\ref{figd}.
We do this calculation primarily to benchmark our numerical method by making
contact with a well-know result from the literature.~\cite{sttern84}

The results of Fig.~\ref{figd} are obtained by solving the following 
coupled equations that constitute the self-consistent Schroedinger-Poisson 
equations including exchange-correlation effects
\begin{equation}
{\frac{-\hbar^2}{2}} {\frac{d}{dz}} \Biggl (      
{\frac {1}{m(z)}} {\frac {d \psi_i(z)}{dz}}
\Biggr ) + V(z) \psi_i(z)= E_i \psi_i (z)
\label{schroedinger}
\end{equation}
\begin{equation}
{\frac {d}{dz}} \Biggl (
\epsilon_o \kappa (z) {\frac {d\phi(z)}{dz}}
\Biggr )
= e \sum_i n_i |\psi_i(z) |^2 - \rho(z)
\label{poisson}
\end{equation}
where $\kappa (z)$ and $\rho(z)$ are the fixed spatially dependent
dielectric function and background charge density of the the interface 
as described in Ref.~\onlinecite{sttern84}.  The potential energy 
of the 2DEG is given by:
\begin{equation}
V(z) = -e \phi(z) + V_{xc}(z)
\label{poten}
\end{equation}
where $\phi(z)$ is the self-consistent potential and $V_{xc}(z)$
is the exchange-correlation potential.  In the above equations, the 
coordinate $z$ is measured relative to the interface between AlGaAs
and GaAs.

The results of Fig.~\ref{figd} are in excellent agreement with previous
results~\cite{sttern84} confirming the soundness of our approach.  

\begin{figure}
\includegraphics[height=6cm]{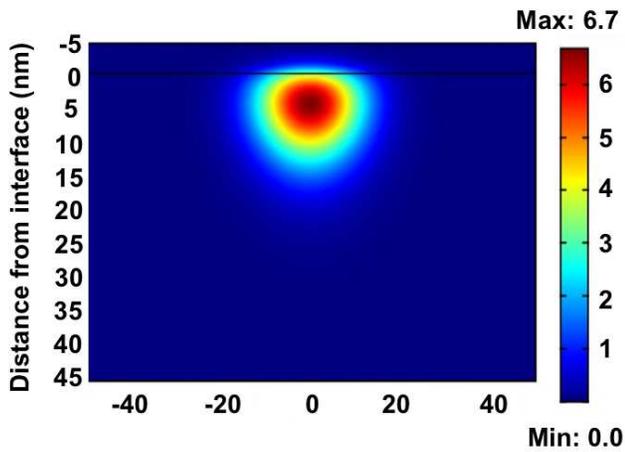}
\caption{\label{fige}(color online) Quantum dot wave function plotted in the 
x-z plane formed by applying a quadratic confining potential in the plane 
(i.e. along the $x$ axis). This potential is characterized by the parameter
$\ell_0 = 20nm$ (quantum dot radius, see Eq.~\ref{elldef}).}
\end{figure}

We next consider the formation of an electrostatically defined quantum dot
by applying a symmetric, confining potential in the plane of the 2DEG 
as illustrated in Fig.~\ref{fige}.  In other words, we add a potential 
of the form 
\begin{equation}
V_x(x) = {\frac {1}{2}} m \omega_o^2 x^2
\end{equation}
to Eq.~\ref{poten} and solve the system of 
Eqs.~\ref{schroedinger},~\ref{poisson} and~\ref{poten} self-consistently
in the two-dimensional $x-z$ domain.  Figure~\ref{fige} clearly illustrates
the formation of a quantum dot in the potential well of the 2DEG as 
expected.

\begin{figure}
\includegraphics[height=6cm]{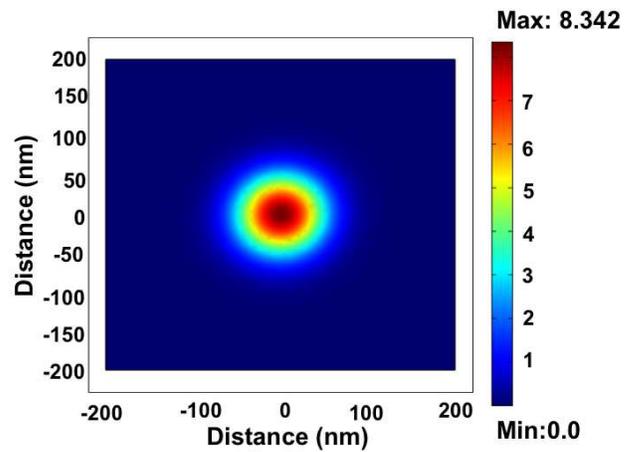}
\caption{\label{figf}(color online) In-plane wave-function for quantum dot 
formed by a symmetric quadratic confining potential (Eq.~\ref{quadpot}) in 
the ($x-y$) plane with $\ell_o = 40nm$ (see Eq.~\ref{elldef}).}
\end{figure}

\begin{figure}
\includegraphics[height=6cm]{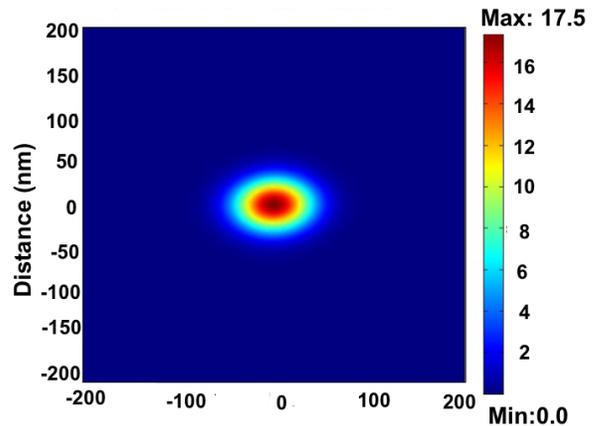}
\caption{\label{figi}(color online) In-plane wave-function for quantum 
dot formed by an asymmetric quadratic confining potential in the ($x-y$) plane.
The parameters for this potential where chosen to mimic the realistic
potential of Figs.~\ref{figb} and~\ref{figc} respectively and are
given by $\ell_o = 30nm$ (see Eq.~\ref{elldef}) and $\alpha = 1$ and 
$\beta = 2.8$ (see Eq.~\ref{quadpot}).
}
\end{figure}

\begin{figure}
\includegraphics[height=6cm]{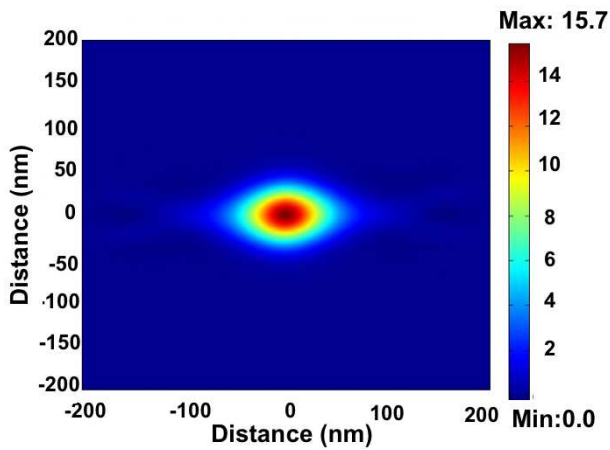}
\caption{\label{figl}(color online) In-plane wave-function for quantum dot 
formed by the potential of Eq.~\ref{separable} illustrated in 
Figs.~\ref{figb} and~\ref{figc}.}
\end{figure}

In the remainder of this paper we focus our attention on motion in the 
plane of the 2DEG and contrast effects associated with quantum dots in 
symmetric and asymmetric confining potentials as illustrated in 
Figs.~\ref{figf} and~\ref{figi}, respectively.  These figures were obtained
using the quadratic model potential of Eq.~\ref{quadpot} with 
$\alpha = \beta = 1$ and $\alpha = 1$; $\beta = 2.8$ respectively in 
the two-dimensional in-plane (i.e. $x-y$ plane) Hamiltonian, $H_{xy}\,$, to 
be discussed in the following.  The parameters of the asymmetric potential
were chosen so as to mimic the realistic potential of Figs.~\ref{figb} 
and~\ref{figc}.  The wave function in the asymmetric model potential
of Fig.~\ref{figi} should be contrasted with the wave function in the 
realistic potential (i.e. using the form of Eq.~\ref{separable})
as shown in Fig.~\ref{figl}.

We consider the motion of the electron in the $x-y$ plane of the quantum
dot in the presence of a magnetic field oriented perpendicular to the 
plane of the 2DEG.  Our approach closely follows that of 
Ref.~\onlinecite{dassarma03}. Thus the total Hamiltonian can be written as
\begin{equation}
H = H_{xy} + H_z + H_{so}
\label{total}
\end{equation}
where $H_z$ corresponds to motion normal to the interface (as discussed
in the context of Eqs.~\ref{schroedinger},~\ref{poisson} and~\ref{poten}),
$H_{so}$ is the spin-orbit interaction to be discussed shortly and the 
remaining term is given by:
\begin{equation}
H_{xy} = {\frac {\vec{P}^2}{2m}} + {\frac{1}{2}} m \omega_o^2 
(\alpha x^2 + \beta y^2) + g_o \mu_B \sigma_z B,
\label{hxy}
\end{equation}
where the kinetic momentum operator:
\begin{equation}
\vec{P} \equiv \vec{p} + {\frac {e}{c}} \vec{A},
\end{equation}
is the sum of the canonical momentum
\begin{equation}
\vec{p} \equiv -i\hbar (\partial_x,\partial_y,0),
\end{equation}
and the vector potential (in the symmetric gauge)
\begin{equation}
\vec{A} \equiv {\frac {B}{2}} (-y,x,0).
\end{equation}

The eigenstates of $H_{xy}$ (Eq.~\ref{hxy}) with $\alpha = \beta$
are the well-known Fock-Darwin states.~\cite{fock28,darwin30}  
The situation with 
$\alpha \ne \beta$ also has an analytic solution.~\cite{schuh85}
We have verified that our numerical solution of $H_{xy} |\psi> = 
\epsilon |\psi>$ is consistent with these analytical results.

Lastly we consider the spin orbit interaction as embodied in the 
Hamiltonian $H_{so}$ which is the essential ingredient in the phenomena
of electric field induced spin switching.~\cite{dassarma03,zutic04}  We write:
\begin{equation}
H_{so} = H_R + H_{D1} + H_{D2}
\end{equation}
where the {\em Rashba} interaction~\cite{rashba60,bychkov84} is given by:
\begin{equation}
H_R = {\frac {\alpha_R e E}{\hbar}} 
\Biggl (
\sigma_x P_y - \sigma_y P_x
\Biggr ),
\label{rashba}
\end{equation}
and the linear and cubic {\em Dresselhous} 
interactions~\cite{dresselhaus55,dyakonov84} are written as:
\begin{equation}
H_{D1} = {\frac {0.7794 \gamma_c k^2 }{\hbar}}
\Biggl (
-\sigma_x P_x + \sigma_y P_y
\Biggr ),
\end{equation}
which is linear in components of the momentum operator $\vec P$ and 
\begin{equation}
H_{D2} = {\frac {\gamma_c}{\hbar^3}}
\Biggl (
-\sigma_x P_x P_y^2 - \sigma_y P_y P_x^2
\Biggr ) + h.c.,
\label{dresselhous2}
\end{equation}
which is cubic in components of the momentum operator ($h.c.$ denotes
the Hermitian conjugate).~\cite{dassarma03}  Note, the electric
field strength $E$ that enters Eq.~\ref{rashba} is that associated with 
the heterojunction $|E| = {\partial V(z)}/{\partial z}$ and is treated
as an adjustable parameter.  Physically we can implement changes in
$E$ through the application of appropriate gate potentials.
All numerical parameters in the above pieces of $H_{so}$ are those
for GaAs found in Ref.~\onlinecite{dassarma03}

The eigenvalue equation $H\, |\,\psi\!> \,=\, \epsilon |\,\psi \!> $, 
with $H$ given by Eqs.~\ref{total} through~\ref{dresselhous2}, was solved 
numerically
to obtain the lowest few eigenstates and eigenenergies vs. the various
parameters of the system.  These parameters include the magnetic field
strength $B$, the electric field $E$, and the strength of the quantum
dot confinement potential.  This latter parameter is conveniently 
characterized by defining the {\em quantum dot radius} as follows:
\begin{equation}
\ell_0 \equiv \sqrt{\frac{\hbar}{m\omega_o}}
\label{elldef}
\end{equation}

The notion of electric field induced spin switching is quantified
by defining an effective electron $g$ factor by the following definition:
\begin{equation}
\epsilon = {\frac {1}{2}} g \mu_B \sigma_z B
\end{equation}
to describe the energy difference between the lowest energy up and 
down spin states. Thus we consider the lowest two states (including 
spin) $\epsilon_2$ and $\epsilon_1$ and calculate the effective $g$ factor 
as:
\begin{equation}
g = {\frac {(\epsilon_2 - \epsilon_1)} {\mu_B B}}.
\end{equation}
Results for the variation of this effective $g$ factor as a function of the
parameters $E$, $B$ and $\ell_o$ is presented in the following section.

%\begin{equation}
%\end{equation}

\section{Results}

We now turn to a presentation of the key results of this work: the 
tunability of the electron $g$-factor through the application of 
electric and magnetic fields.

\begin{figure}
\includegraphics[height=6cm]{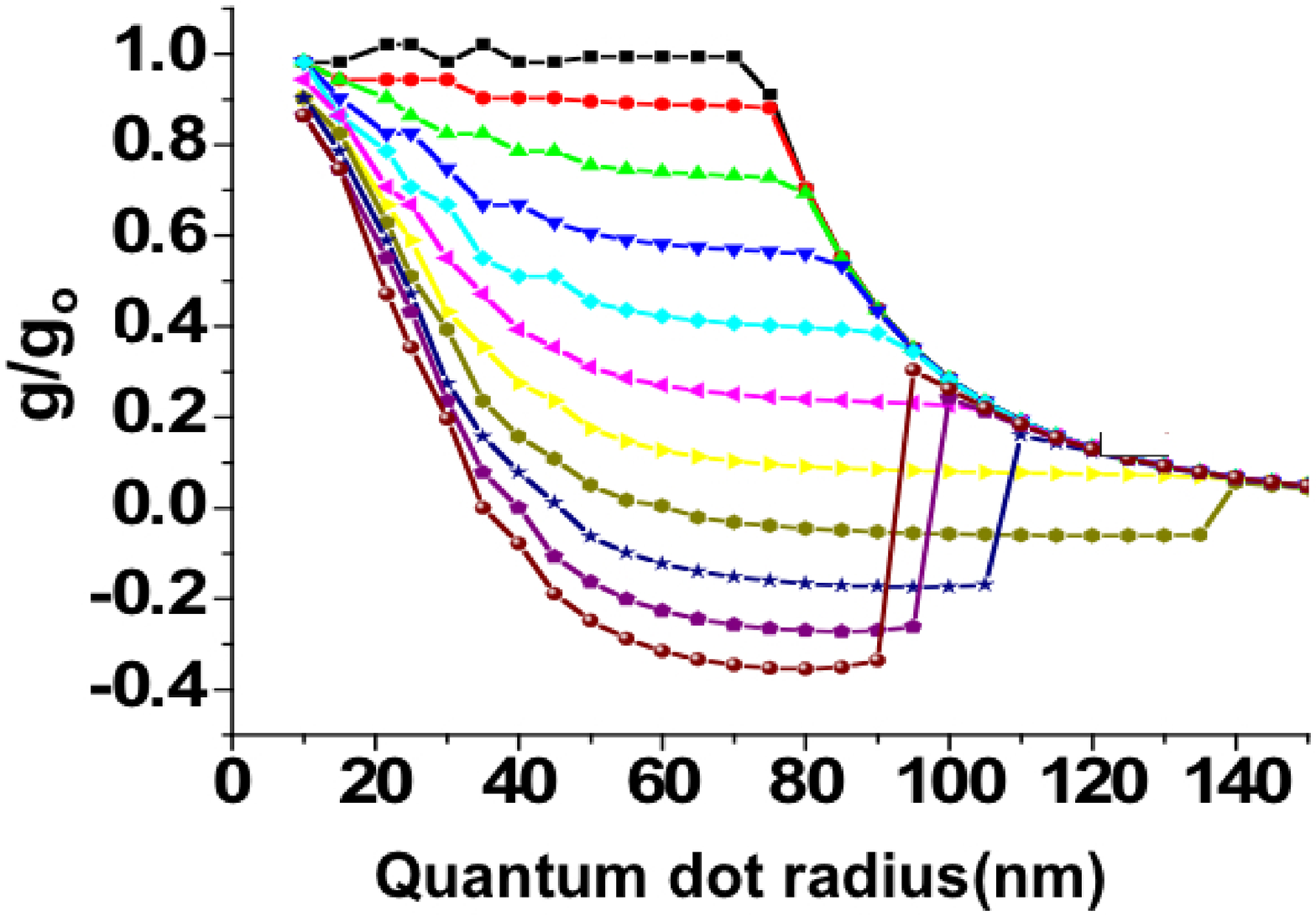}
\caption{\label{figg}(color online) Electric field induced changes in the 
$g$-factor vs. quantum dot radius for various electric field strengths for 
a symmetric quantum dot in the quadratic potential of Eq.~\ref{quadpot}.  From 
top to bottom, the 
curves represent increasing electric field strength as follows.  The first
curve corresponds to $1\times 10^4 V/cm$, and the rest range from
$1\times 10^5 V/cm$ through $1\times 10^6 V/cm$ in equal steps with $B = 1T$.
This results is consistent with Ref.~\onlinecite{dassarma03}.}
\end{figure}

Figure~\ref{figg} is consistent with previous published work~\cite{dassarma03} 
and 
illustrates the $g$-factor tunability vs. the strength of the applied 
electric field and confining potential (as parametrized by the 
quantum dot radius $\ell_o$) for fixed magnetic field ($B = 1T$) for the 
symmetric quantum dot in the quadratic potential of Eq.~\ref{quadpot}
with $\alpha = \beta = 1$.

\begin{figure}
\includegraphics[height=6cm]{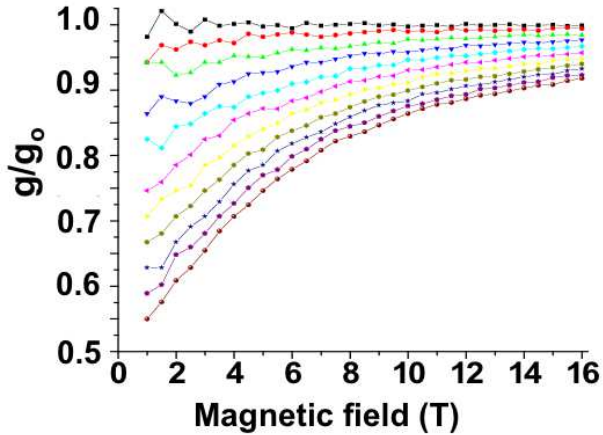}
\caption{\label{figh}(color online) Electric field induced changes in the 
$g$-factor vs. magnetic field for various electric field strengths for the 
symmetric quantum dot.  From top to bottom, the 
curves represent increasing electric field strength as follows.  The first
curve corresponds to $1\times 10^4 V/cm$, and the rest range from
$1\times 10^5 V/cm$ through $1\times 10^6 V/cm$ in equal steps. For this 
calculation, the quantum dot radius was fixed at $\ell_o = 20nm$}
\end{figure}

Figure~\ref{figh} is also consistent with previous published 
work~\cite{dassarma03} and 
illustrates the $g$-factor tunability vs. the strength of the applied 
electric field and and magnetic field for fixed confining potential
(parametrized by the quantum dot radius $\ell_o = 20nm$).

\begin{figure}
\includegraphics[height=6cm]{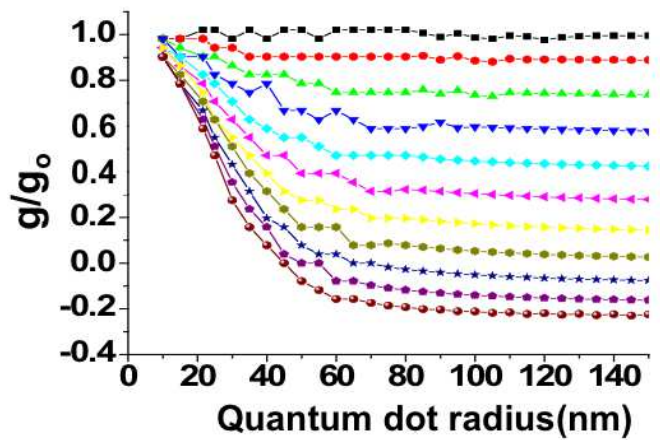}
\caption{\label{figj}(color online) Electric field induced changes in the 
$g$-factor vs. quantum dot radius for various electric field strengths for 
an asymmetric quantum dot (wave function not shown).  From top to bottom, the 
curves represent increasing electric field strength as follows.  The first
curve corresponds to $1\times 10^4 V/cm$, and the rest range from
$1\times 10^5 V/cm$ through $1\times 10^6 V/cm$ in equal steps. Again
we choose $B = 1T$.}
\end{figure}

Upon introducing in-plane anisotropy to the confining potential we find 
significant changes in the electric-field induced $g$-factor tunability
as illustrated in Fig.~\ref{figj}.  This figure was generated by choosing
$\alpha = 1$, $\beta = 2$ and $\ell_o = 120nm$.  
%The resulting wave function
%was discussed earlier and was illustrated in Fig.~\ref{figi}.  

\begin{figure}
\includegraphics[height=5cm]{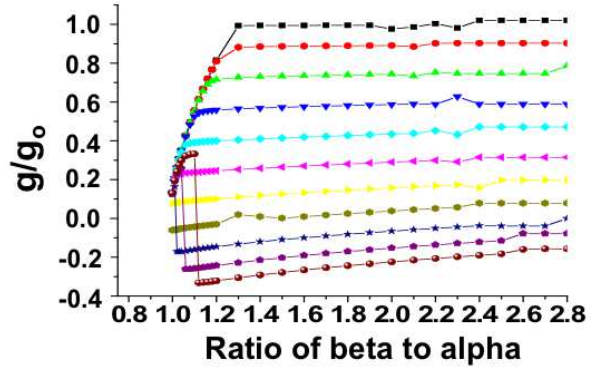}
\caption{\label{figk}(color online) Electic field induced changes in the 
$g$-factor vs. the degree of anisotropy of the quantum dot confinement 
potential for various electric field strengths.  From top to bottom, the 
curves represent increasing electric field strength as follows.  The first
curve corresponds to $1\times 10^4 V/cm$, and the rest range from
$1\times 10^5 V/cm$ through $1\times 10^6 V/cm$ in equal steps.}
\end{figure}

To quantify the effects of in-plane anisotropy, we have carried out a 
parameter study of the $g$-factor tunability vs. degree of anisotropy and 
the results are presented in Fig.~\ref{figk}.  This figure was generated
by fixing the quantum dot radius at $\ell_o = 120nm$ and holding $\alpha = 1$
while varying $\beta$ with $B = 1T$.

\begin{figure}
\includegraphics[height=6cm]{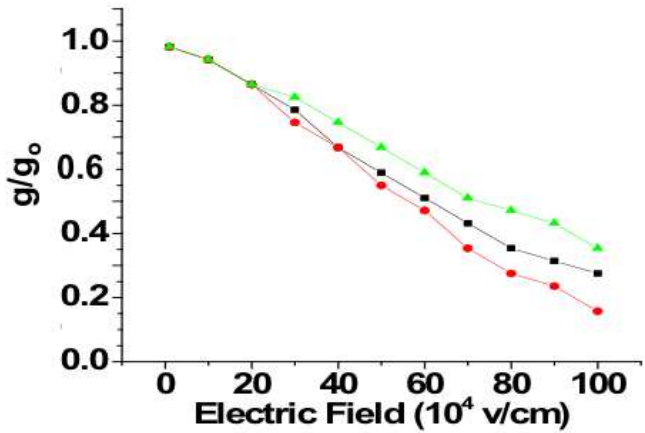}
\caption{\label{fign}(color online) Changes in the 
$g$-factor vs. electric field for quantum dots in the realistic 
potential of Figs.~\ref{figb} and~\ref{figc} (using the 
form of Eq.~\ref{separable}) in comparison to results for symmetric
and asymmetric model potentials.  The black (middle) curve corresponds to the 
realistic potential while the red (lower) and green (upper) curves correspond 
to symmetric and asymmetric model potentials respectively.  The parameters
were chosen to closely mimic the realistic potential and are given by
$\ell_o = 30nm$ and $\alpha = \beta = 1$ for the symmetric potential and 
$\alpha = 1$; $\beta = 2.8$ for the asymmetric potential. }
\end{figure}

Lastly we consider the results for $g$-factor tunability for quantum 
dots in the realistic potential of Figs.~\ref{figb} and~\ref{figc} and
Eq.~\ref{separable}. Figure~\ref{fign} illustrates the results for 
a quantum dot in the realistic potential (middle curve) in comparison with 
symmetric (lower curve) and asymmetric (upper curve) quantum dots.
The parameters were chosen to closely mimic the realistic potential and are 
given by $\ell_o = 30nm$ and $\alpha = \beta = 1$ for the symmetric potential 
and $\alpha = 1$; $\beta = 2.8$ for the asymmetric potential.  

From Fig.~\ref{fign} we conclude that the realistic potential result
is bracketed by those for symmetric and asymmetric model potentials and,
as such, contains characteristics of each.

\section{Discussion}

The key result of this work is illustrated in Figs.~\ref{figg} 
and~\ref{figj}: anisotropy in the confining potential significantly 
extends the size range of quantum dots that exhibit electric field induced 
$g$-factor tunability.  Indeed, in Fig.~\ref{figg} we see that all of the 
curves collapse onto a single curve for large quantum dots (i.e. larger 
than $\ell_o = 120nm$) negating the switching effect.  With anisotropy, 
however, there is no degradation of the switching effect for large dots.

Another important result of this work is the realization that the degree
of anisotropy need not be very large in order to obtain significant changes 
in the gate induced $g$-factor tunability, as illustrated in Fig.~\ref{figk}.
We see that the maximum effect is obtained for roughly 
$\beta/\alpha \approx 1.3$ and begins to decrease slightly for larger values
of this shape-anisotropy ratio.  The jumps in the value of the $g$-factor
from positive to negative value are indicative of level crossings 
(e.g. the relative ordering of spin up and down levels changes as a function
of the anisotropy).

Lastly we have seen from Fig.~\ref{fign} that results for quantum dots
in realistic potentials contain characteristics of both the symmetric
and asymmetric model potentials.  This emphasizes the need for realistic
numerical simulations since the physics of such systems cannot be fully
captured by symmetric or asymmetric model potentials alone.

\section{Conclusions}

We have carried out a numerical simulation study of gate induced 
tunability of the electron $g$-factor in a prototype single electron 
spintronic device.  We consider a realistic three-dimensional geometry
and employ a numerical approach based on the finite element 
method~\cite{comsol}.  

Due to the large disparity in physical length scales we have adopted an 
approximate strategy as a complete self-consistent solution of the full 
problem is prohibitive given existing computational tools.  In our approach 
we have investigated the problem in stages.  

%In the first step we consider 
%the electrostatic problem of our device under the assumption that the 2DEG 
%is approximated a classical perfect conductor.  This allows us to determine
%the spatial variation of the potential in the plane of 2DEG and we note
%that it is anisotropic.  This result is the motivation for studying
%effects of anistropy of the quantum dot confining potential.

%In the second step we have numerically studied the formation of the 
%2DEG at the heterojunction between AlGaAs and GaAs by solving for the 
%wave functions and self-consistent potential.  Our results are in excellent
%agreement with previous work.  We also consider the formation of a 
%quantum dot by applying a quadratic (harmonic oscillator) potential to the 
%Hamiltonian describing the 2DEG formation.  The quantum dot formation is 
%observed as expected.  These facts provides a useful benchmark validation of 
%the approach.

%Finally we turn our focus to the properties of the quantum dot formed in
%2DEG and investigate the electric field tunability of the effective
%$g$ factor vs. applied electric fields, magnetic fields and confinement
%potentials. 

We find that symmetry breaking due to anisotropy of the quantum dot
confining potential leads to significant changes in the gate induced 
tunability of the $g$-factor vs. quantum dot radius extending the size
range of quantum dots that can be tuned.  We also find that the anisotropy
ratio need only be on the order of $\beta/\alpha = 1.3$ to produce the 
largest effects.

By employing non-perturbative, fully numerical methods and realistic 
geometries, our approach is providing insights that might be difficult 
or impossible to obtain using analytical techniques alone.

\bibliography{paper}

\end{document}